\documentclass[12pt, a4paper]{article}

\usepackage{graphicx}
\usepackage[hidelinks]{hyperref}
\usepackage{xcolor}
\usepackage{authblk}  
\usepackage[margin=1in]{geometry}  

\usepackage[numbers,sort&compress]{natbib}  
\title{Teaching RDM in a smart advanced inorganic lab course and its provision in the DALIA platform}

\author[1]{Alexander Hoffmann}
\author[1]{Jochen Ortmeyer}
\author[2]{Fabian Fink}
\author[1]{Charles Tapley Hoyt}
\author[3]{Jonathan D. Geiger}
\author[3]{Paul Kehrein}
\author[3]{Torsten Schrade}
\author[1,*]{Sonja Herres-Pawlis}

\affil[1]{Institute of Inorganic Chemistry, RWTH Aachen University, Landoltweg 1a, 52074 Aachen, Germany}

\affil[2]{University Library, RWTH Aachen University, Templergraben 61, 52062 Aachen, Germany}

\affil[3]{Academy of Sciences and Literature Mainz, Geschwister-Scholl-Straße 2, 55131 Mainz, Germany}

\affil[*]{To whom correspondence should be addressed: sonja.herres-pawlis@ac.rwth-aachen.de}

\date{}

\begin{document}

\maketitle

\abstract{
Research data management (RDM) is a key data literacy skill that chemistry students must acquire. Concepts such as the FAIR data principles (Findable, Accessible, Interoperable, Reusable) should be taught and applied in undergraduate studies already. Traditionally, research data from labs, theses, and internships were handwritten and stored in inaccessible formats such as PDFs, limiting reuse and machine learning applications. At RWTH Aachen University, a fifth-semester lab course introduces students to the electronic laboratory notebook (ELN) Chemotion, an open-source DFG-funded tool linked to the national NFDI4Chem initiative. Students plan, document, and evaluate experiments digitally, ensuring metadata and analysis are captured for long-term reuse. Chemotion’s intuitive interface and repository enable sustainable data sharing. To reinforce RDM, students receive a seminar and access to online training videos with interactive Moodle elements. Herein we highlight the use of the DALIA platform as a discovery tool for the students.
}

\section{Introduction}\label{sec1}

The digitalization of chemistry promises to support reproducibility, promote transparency, and help address some of the great societal and technological challenges of the modern age~\cite{fantke2021}.
Initial progress in cheminformatics~\cite{steinbeck2025}, virtual screening~\cite{yang2024}, laboratory automation~\cite{bai2022,moshantaf2024}, and other chemistry subdomains has crucially relied on the availability of chemical data and knowledge in a structured, machine-actionable form.

While these structured artifacts have historically been organized through the work of manual curation in concert with semi-automated and automated approaches~\cite{burge2012,isb2018}, the emerging era of chemistry research is more driven by digital-first instrumentation, laboratory notebooks, computational workflows, and chemical information systems~\cite{jablonka2022}. This shift motivates chemists at all career stages to learn best practices in research data management (RDM) to support reproducibility, promote transparency, and  enable reuse. These both include named philosophies such as the FAIR (Findable, Accessible, Interoperable, Reusable) principles~\cite{Wilkinson2016}, TRUST (Transparency, Responsibility, User focus, Sustainability and Technology) principles~\cite{Lin2020}, the O3 (Open Data, Open Code, Open Infrastructure) guidelines~\cite{Hoyt2024} as well as numerous soft skills that have typically been underappreciated in the laboratory sciences.

Within NFDI4Chem\footnote{\url{https://www.nfdi4chem.de}}~\cite{nfdi4chem,nfdi4chem2}, the chemistry consortium of the German National Research Data Infrastructure (NFDI), members strive to build a FAIR and open infrastructure for research data across all areas of chemistry. The consortium develops standards, tools, and training materials to enable chemists to manage, share, and reuse their data effectively. For organic chemistry, some digital tools are already available, such as the Chemotion electronic lab notebook (ELN) and associated data repository. The use of these tools in inorganic chemistry has also begun recently~\cite{Fink2022}.

In inorganic chemistry, experiments often generate diverse and complex data (e.g., different spectra, crystal structures, reaction conditions) that should be preserved in standardized, machine-readable formats. Without effective RDM learning, researchers are often not equipped with the skills necessary to organize, document, and share their work. By learning RDM, they not only increase their own research efficacy, but make valuable contributions towards open science, data-driven discovery, and the applications of artificial intelligence and machine learning~\cite{HerresPawlis2022}. Given the variety of general and domain-specific topics and skills needed for effective RDM, many universities have begun to offer RDM coursework and training for doctoral students, post-doctoral researchers, and faculty. These typically cover a combination of generic and domain-specific topics in data organization, backup, archiving, legal compliance, and FAIR data principles through online self-learning modules incorporating video content, interactive modules, and virtual learning platforms.

We believe that RDM should be taught as early as in undergraduate chemistry curricula to give students earlier practical exposure to RDM practices. We have previously demonstrated one of the first applications in the German chemistry education landscape~\cite{Fink2023} and have additional evidence of the efficacy of teaching RDM to undergraduate students from the engineering discipline run by members of the German National Research Data Infrastructure Engineering Consortium (NFDI4ING) entitled "Scientific Experiment with Integrated Research Data Management” in BSc Mechanical Engineering" \cite{Farnbacher2025}.

\begin{figure}[htbp]
\centering
\includegraphics[width=\linewidth]{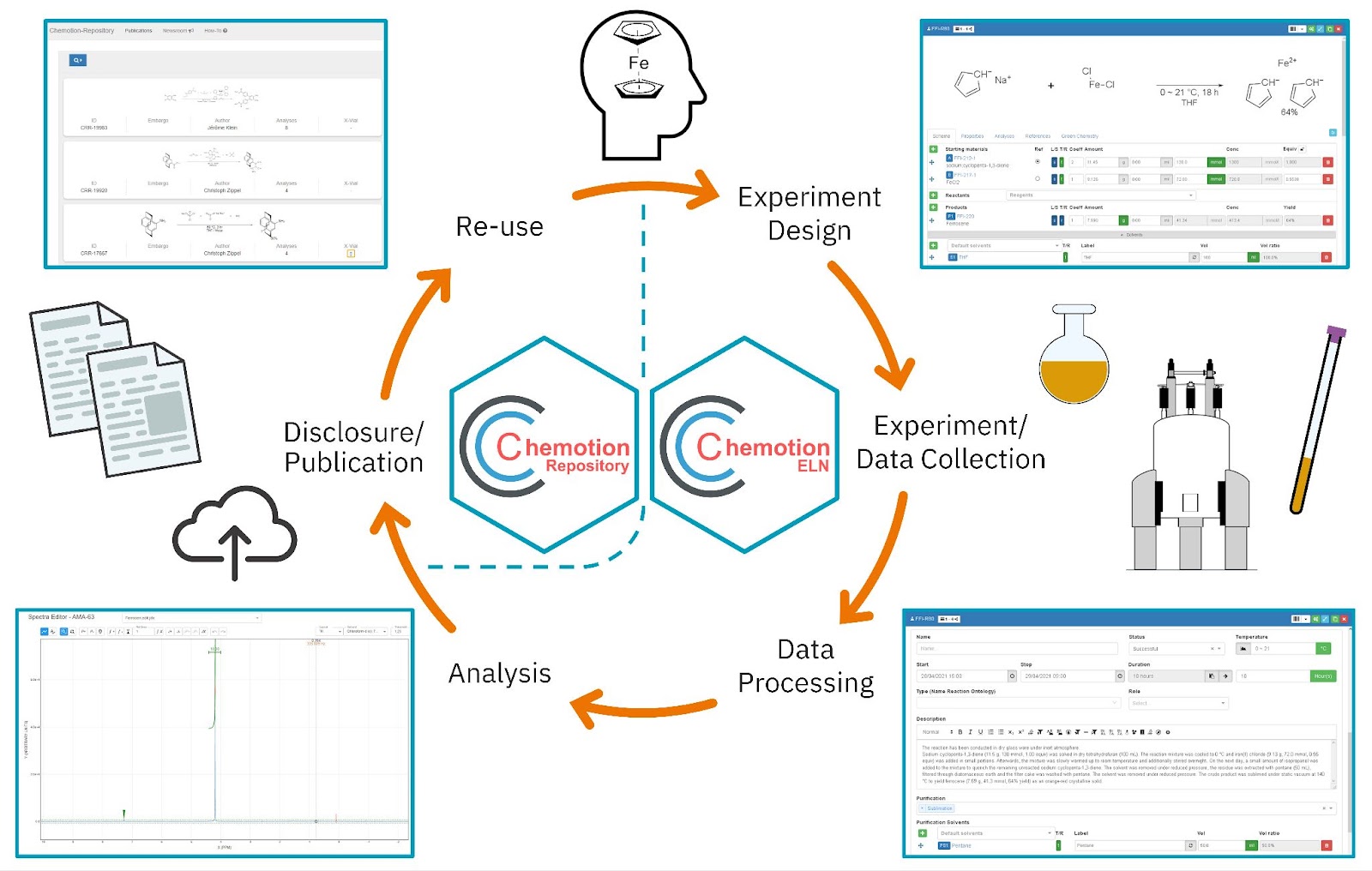}
\caption{Chemotion: Data lifecycle realized in Chemotion as an electronic laboratory notebook and as a repository.}\label{figure-chemotion}
\end{figure}

RDM learners, teachers, and teaching material creators face different issues. Learners often struggle with abstract concepts (e.g., FAIR principles, metadata standards) when they are not directly linked to concrete laboratory practices. Further, chemistry students already face a dense curriculum, so RDM can feel like an additional burden rather than an integrated skill. The wide variety of digital tools including electronic laboratory notebooks (ELNs), repositories, and data standards also creates uncertainty about which practices will be most relevant to their future careers. Moreover, students come with very different levels of digital literacy, which makes it difficult to achieve consistent progress across a group. RDM teachers often face the challenge that they were not themselves trained in RDM and may lack confidence in teaching it. Similarly to learners, limited course time forces teachers to balance experimental work with digital documentation, while evolving standards and infrastructures require constant updating. Evaluating whether students manage data correctly also remains difficult. Finally, creators of RDM teaching materials face the issues of identifying and adapting general, abstract RDM concepts with chemistry-specific use cases, ensuring materials remain up-to-date with the ever-evolving landscape within chemistry and generic best practices. Because generic RDM teaching material is not well-received by undergraduate chemistry learners, hands-on learning presents an alternative strategy to support them. Therefore, we have identified ELNs as an ideal avenue through which to teach RDM to students in their lab stages that addresses issues for learners, teachers, and training material creators. Later, we demonstrate teaching RDM using the Chemotion ELN~\cite{Tremouilhac2017}, an open-source ELN integrated with many chemistry laboratories via NFDI4Chem that supports researchers in digital experiment planning, metadata capture, repository upload, and visualization of the full data life cycle (Figure \ref{figure-chemotion}).

The creation and distribution of reusable teaching materials in order to improve access for learners, better prepare teachers, and reduce the burden on teaching materials creators resulted in the advent of open educational resources (OERs)~\cite{Jung2020}.
Following the publication of new guidelines from Deutsche Forschungsgemeinschaft (DFG) in 2010~\cite{DFG2010} and founding of the Research Data Alliance (RDA)~\cite{RDA2014}, increasing focus has been placed on the development of OERs for RDM as funders and universities increasingly required data management plans and transparent research practices. Early initiatives such as MANTRA\footnote{\url{https://mantra.ed.ac.uk}} and DataONE Education Modules\footnote{\url{https://www.dataone.org/training}} provided freely available training materials, often as self-paced online courses. The UK Digital Curation Centre\footnote{\url{https://www.dcc.ac.uk}} developed widely adopted guidance and checklists, which were shared openly and became foundational references. With the rise of the FAIR principles~\cite{Wilkinson2016}, international projects expanded the scope of OERs beyond compliance to emphasize interoperability and reuse. Programs like FOSTER Open Science\footnote{\url{https://openscience.eu/article/project/foster}} and FAIRsFAIR\footnote{\url{https://www.fairsfair.eu}} created multilingual, discipline-neutral training packages under open licenses~\cite{Kersloot2025}. Focus shifted in the 2020s to discipline-specific OERs as generic OERs were demonstrated to have shortcomings in impacting the culture of research and teaching~\cite{ComparisonRDI}. National and international initiatives such as NFDI4Chem, NFDI4Ing, and NFDI4Culture in Germany; Physical Sciences Data Infrastructure (PSDI)\footnote{\url{https://www.psdi.ac.uk}} in the UK; and ELIXIR across Europe have developed tailored tutorials, videos, and lab course integrations as OERs. 
Such initiatives distribute their teaching materials through a variety of venues.
For example, PSDI makes its teaching materials available via Moodle\footnote{\url{https://resources.psdi.ac.uk/service/62af1a1c-f4c9-48ee-acc2-45b19fb8b231}}.
Some initiatives have adopted generalist repositories like Zenodo and Figshare that make teaching materials more FAIR and accessible by assigning persistent identifiers and annotating to them structured metadata.
Others have deployed custom institutional repositories or repurposed external systems like GitHub for distribution.
Today, the development of OERs for RDM has become a global, collaborative effort supported by networks such as the RDA and the GO FAIR initiative\footnote{\url{https://www.go-fair.org}}, which encourage sharing, adaptation, and co-creation of training materials across disciplines.

Perifanou and Economides (2023) semi-automatically identified and characterized nearly six thousand unique OER repositories covering a variety of media types (e.g., articles, books) and scopes (e.g., region-specific, domain-specific, institution-specific, and generic)~\cite{Perifanou2023}. The variety of metadata standards and implementations in the OER repository landscape led to the development of several reusable data models/schemata and OER platforms. For example, the Dublin Core Metadata Initiative's Learning Resource Metadata Innovation (LMRI) and Educational Resource Discovery Index~\cite{erudite} groups contributed an educational resource schema to Schema.org~\cite{Guha2016}, the W3 Open Educational Resources Schema Community Group developed the OERSchema\footnote{\url{https://github.com/open-curriculum/oerschema}}, and many isolated efforts like ProvOER~\cite{RibeiroDosSantos2023} developed ancillary metadata models. In this context, Oellers and Rörtgen provided a very helpful compendium on didactic metadata~\cite{Oellers2024}. Several ontologies and controlled vocabularies were generated or extended to support data modeling in these fields, which we reviewed and contributed as a collection (\url{https://bioregistry.io/collection/0000018}) to the Bioregistry, a comprehensive registry of ontologies and controlled vocabularies~\cite{Hoyt2022}. The ELIXIR Training e-Support System (TeSS)~\cite{Beard2020} made an early demonstration in the biomedical domain of an open source, reusable software platform that both indexes existing OER platforms in a harmonized data model as well as enables the contribution of new OERs. However, TeSS lacked several key metadata fields supporting learners in chemistry and was difficult to extend, motivating the development of an alternative platform to best support RDM learners and chemists.

In this article, we present four major contributions in teaching RDM to chemists. First, we performed a comprehensive survey of open educational resources for chemistry, highlighting materials in RDM. Second, we developed training material to fill existing gaps and specifically support bachelor's, masters', and doctoral students in using ELNs as an ideal tool for RDM. Third, we share lessons learned and best practices for developing training materials for chemists, making them available as open educational resources, and reaching as a prototype for teaching digital chemistry to the community. Finally, we present the DALIA OER Platform (\url{https://dalia.education/en}) for indexing open educational resources and making them best accessible to students.

\section{Results}\label{results}

To achieve an integration of RDM into science and to foster a cultural change, it is necessary to include the topic into the curricular teaching of undergraduate students. With this, young scientists engage with the topic at the earliest stage of their academic career already and can apply the competences acquired in their future studies and later on in industry. In natural sciences, a promising approach is the integration of ELN software in practical laboratory courses as a hands-on experience. To ensure the success of this approach, the ELN integration has to be accompanied by fitting learning materials on RDM in general.

There are many generic RDM learning materials available which are suitable even for undergraduate students\footnote{\url{https://www.rwth-aachen.de/cms/root/Forschung/Forschungsdatenmanagement/Weiterbildungsangebote/~udzt/Lehrvideos/}}. We observed that using such generic materials in a lecture series we hosted entitled sustainable coordinative polymerization catalysis led to a low engagement with and understanding of the RDM content among the chemistry students. This led to the conclusion that more lifelike, discipline-specific materials were necessary. Consequently, we developed chemistry-specific RDM teaching materials, which are more tangible for students so they can better relate to the content. The material comprises teaching videos\footnote{\url{https://av.tib.eu/series/1527}} based on general RDM learning materials\footnote{\url{https://www.youtube.com/playlist?list=PL0eaiqVqG1ovx-A\_rGpz76nylfLjd9auR}}~\cite{Kraft2024} and interactive Moodle elements with a final test on RDM~\cite{Fink2023}. Since practical lab states are evolving to new experiments, the set of videos will be extended continuously.

\subsection{A Case Study with Chemotion}\label{case-study-chemotion}

For the successful integration of the Chemotion ELN in a fifth-semester advanced inorganic lab course at RWTH Aachen University, teaching videos on the use of the ELN were recorded and provided to the students helping them to get started using the software. These instruction videos are based on the online documentation guide of the ELN\footnote{\url{https://chemotion.net/docs}} and cover its basic functions supplemented with specific information for the students of the lab course. Thus, the videos complement the written online documentation and offer an audiovisual learning approach as an alternative. A survey among the students of the lab course revealed a general satisfaction with and acceptance of the provided instruction videos~\cite{Fink2023}. The full set of videos was published on Zenodo~\cite{Fink2023b} and made available via YouTube\footnote{\url{https://www.youtube.com/playlist?list=PL1AonKd9WAd8cDjzXGiNu0ndoctNu0izs}}. The German and English versions of the videos have been downloaded more than 1,600 times from Zenodo and have more than 5,000 views on YouTube in total showing their broad impact on the community. Additionally, this OER is indexed in the DALIA platform under record \url{https://search.dalia.education/items/b37ddf6e-f136-4230-8418-faf18c4c34d2} (Figure \ref{figure-dalia}, right). Here, all relevant metadata of the Zenodo publication, e.g., authors, description, license, and keywords, are displayed together with further information, e.g., on the community, discipline, target group, and proficiency level. DALIA’s convenient search function allows for an easy and straightforward discovery of OERs based on the mentioned metadata and information further enhancing the video’s impact.

\subsection{RDM Landscape in DALIA}\label{rdm-landscape-dalia}

The landscape of RDM training materials is currently fragmented across OER platforms and repositories that are typically specific for a region, institution, or project. Further, existing OER platforms and repositories typically do not annotate OERs with important metadata that promotes findability to learners such as the resource type (e.g., lecture, tutorial, podcast), media type (e.g., text, audio, video), discipline (e.g., chemistry), target groups (e.g., bachelor's, master's, doctorate), and learner proficiency level (e.g., novice, intermediate, advanced). Therefore, we developed the DALIA OER Platform (\url{https://dalia.education/en}; see methods) to address these needs and provide a portal that both indexes the OERs developed for our use case as well as aggregating external OERs for RDM. We identified these external OERs through discussions with stakeholders in the NFDI4Chem Consortium including principle investigators and wet-laboratory researchers. 

\begin{figure}[htbp]
\centering
\includegraphics[width=\linewidth]{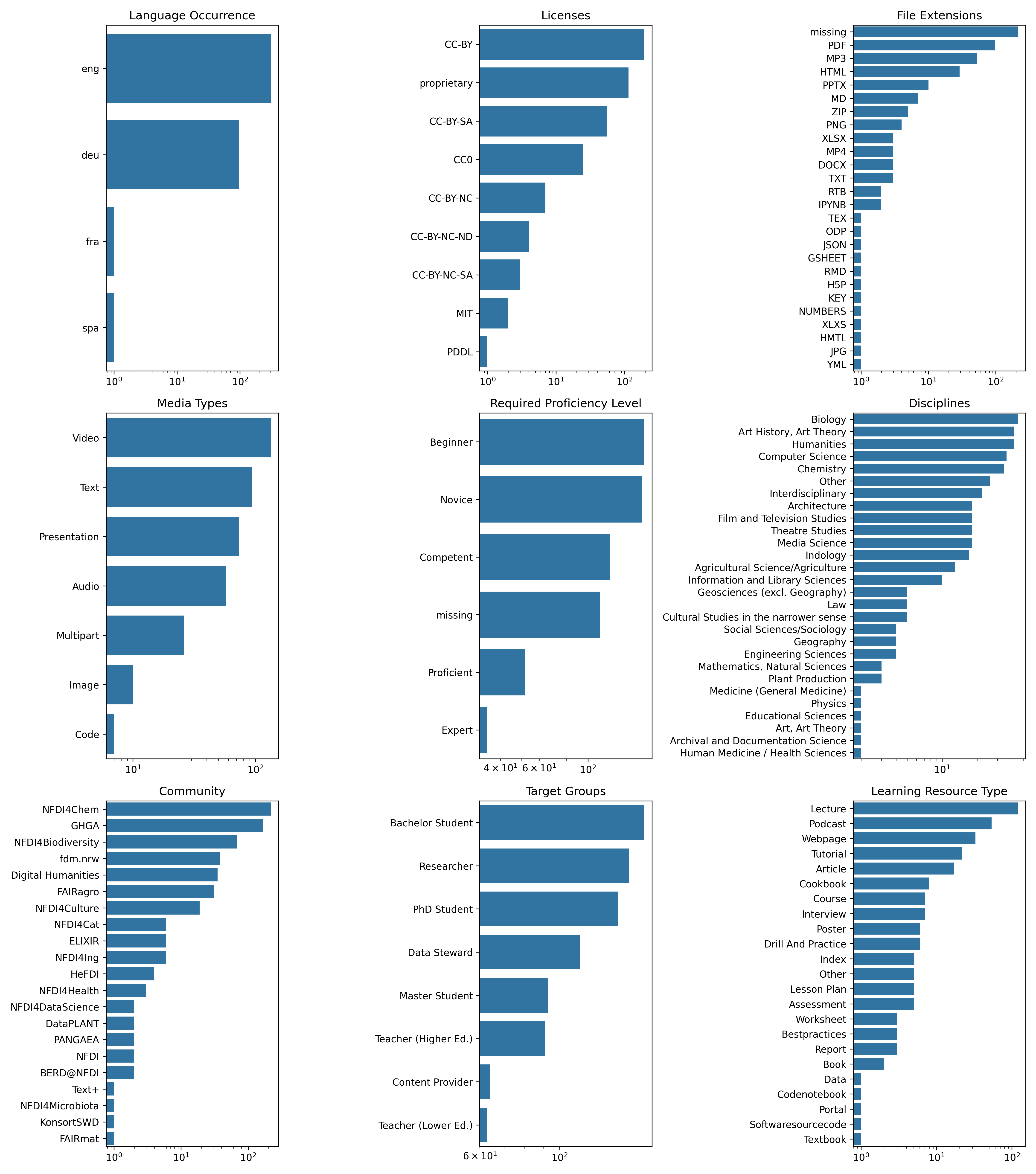}
\caption{A summary of the open educational resources (OERs) in the DALIA OER Platform.}\label{figure-summary}
\end{figure}

Ultimately, DALIA indexes dozens of OERs in the chemistry RDM space and hundreds of OERs in other disciplines and topic areas for a total of 405 OERs (Figure \ref{figure-summary}). These resources cover a variety of target groups, resource types, and learner proficiency levels to make learning accessible to students and other researchers with different backgrounds and learning styles. Notably, DALIA has the largest number of OERs targeted towards undergraduate (bachelor) students (172/405; 42.5\%) compared to other target groups and beginners (180/405; 44.4\%) compared to other required proficiency levels.

However, we note that the annotation of OERs with high-quality metadata required to support findability is difficult, time-consuming, and requires the careful development of both internal and external controlled vocabularies and data models. Therefore, we aspire towards collaboratively improving these controlled vocabularies and data models and continually improving the metadata annotations to maintain DALIA as a FAIR, sustainable, and up-to-date resource.

\section{Discussion}\label{discussion}

The integration of RDM into undergraduate chemistry curricula through ELNs and tailored OERs has demonstrated clear benefits in student engagement and skill development. Our experience with the Chemotion ELN in an advanced inorganic laboratory course shows that coupling hands-on experimental work with digital documentation enhances student understanding of the FAIR principles and makes abstract concepts such as metadata, interoperability, and data reuse more tangible. Compared to the use of generic RDM teaching materials, the discipline-specific approach yielded stronger engagement, which underlines the importance of contextualization in data literacy education.

At the same time, our findings highlight several gaps that motivate further pedagogical and technical development. From a chemistry perspective, future work must extend beyond inorganic chemistry to encompass other subdisciplines such as organic and physical chemistry, where different data types and workflows create new challenges. Moreover, while DALIA successfully enhances the findability and accessibility of OERs, our metadata analysis revealed persistent inconsistencies in descriptors such as discipline, keywords, and proficiency level. These gaps limit precision in search and reuse, emphasizing the need for more standardized vocabularies, semi-automated metadata enrichment, and community-driven curation practices.

Another point of discussion concerns the transferability of our approach. Many of the principles and tools described here are not unique to chemistry but could be extended to other experimental sciences such as physics, biology, or engineering. For example, the successful integration of Jupyter-based training in physics curricula suggests opportunities for cross-disciplinary exchange of OERs via platforms like DALIA. Beyond the natural sciences, the modularity of OER and the universality of the FAIR, TRUST, and O3 principles suggest that similar approaches may benefit fields further removed from laboratory practice, such as the social sciences or humanities, albeit with discipline-specific adaptations.

Looking forward, the main challenge lies in scaling up while maintaining quality. DALIA’s strength as a domain-sensitive discovery platform must be combined with workflows that allow the semi-automatic harvesting of resources from generic repositories such as Zenodo or DARIAH-Campus\footnote{\url{https://campus.dariah.eu/}}, while still preserving rich, discipline-specific metadata. Pedagogically, a key goal will be to design coherent learning paths that integrate RDM and digital literacy across the entire chemistry curriculum rather than leaving them isolated to individual courses. This requires coordination across teaching staff, alignment with institutional policies, and continued engagement with national and international infrastructures such as NFDI4Chem.

In summary, the combination of discipline-specific RDM training, OER publication, and integration into DALIA constitutes a promising model for embedding data literacy in undergraduate chemistry education. The case study presented here not only demonstrates immediate benefits for student learning but also points toward broader cultural change in research practices. Continued development of technical infrastructures and pedagogical frameworks will be essential to realize the full potential of digital and open science training for the next generation of scientists.

\section{Methods}\label{methods}

We highlight two parts to our methods. First, we describe how we developed open educational resources (OERs) for research data management (RDM). Second, we describe the development and usage of the DALIA OER platform.

\subsection{Developing RDM OERs}\label{rdm-oer-development}

The creation of RDM OERs requires a systematic approach that goes beyond defining learning outcomes. A central starting point is the learning objectives matrix~\cite{Petersen2025}, which helps align competencies with specific target groups such as students, doctoral researchers, or data stewards. However, additional guidelines highlight further quality dimensions. For instance, the FAIR principles for educational resources emphasize making OERs Findable, Accessible, Interoperable and Reusable by providing rich metadata, persistent identifiers, and open file formats, while ensuring clear licensing and access rules. More general OER guidance stresses the importance of accessibility, modularity, and usability, including the provision of alternative formats and attention to copyright issues~\cite{OpenEd,UnaEuropa}. In the RDM context, these aspects are particularly relevant since many teaching examples involve sensitive data, legal restrictions, or ethical considerations, requiring the OER itself to model good practices of documentation, versioning, and attribution~\cite{OpenScienceTrainingHandbook}. Community-driven initiatives such as NFDI4Ing recommend embedding interactive elements like quizzes or H5P modules and engaging disciplinary stakeholders to ensure relevance and uptake~\cite{Farnbacher2025}. Beyond the didactic structure, the LEARN project toolkit\footnote{\url{https://learn-rdm.eu/en/dissemination/toolkit}} underlines the necessity of including institutional and policy perspectives in training resources, broadening the scope of RDM OERs to governance and infrastructure issues. Taken together, these guidelines point to a multi-layered approach: while the learning objectives matrix ensures competence orientation, FAIR and OER principles ensure technical and legal robustness, and community frameworks secure disciplinary and institutional relevance.

\subsection{DALIA OER Platform}\label{dalia}

We implemented the DALIA OER Platform (\url{https://dalia.education/en}) to enable indexing external OERs with comprehensive, FAIR metadata and to enable search by learners and teachers in order to find and access relevant OERs.

DALIA implements a metadata model for OERs that includes a title, description, language, keywords, license, external URL, date published, and author list, dubbed the DALIA Interchange Format (DIF)~\cite{Geiger2024}. DALIA further classifies OERs based on external controlled vocabularies and ontologies for the resource type (e.g., lecture, tutorial, podcast), media type (e.g., text, audio, video), discipline (e.g., chemistry), target groups (e.g., bachelor's, master's, doctorate), and learner proficiency level (e.g., novice, intermediate, advanced). DALIA also links OERs to relevant communities and collections. The metadata model is implemented using semantic web standards such as the Resource Description Framework (RDF) and reuses the Schema.org educational resource data model to enable interoperability and extends it to support the needs of learners and teachers in the chemistry domain. In addition to its specific metadata fields, DALIA has the benefit over generic repositories like Zenodo and Figshare because content is already pre-filtered for educational/training material.

\begin{figure}[htbp]
\centering
\includegraphics[width=0.49\linewidth]{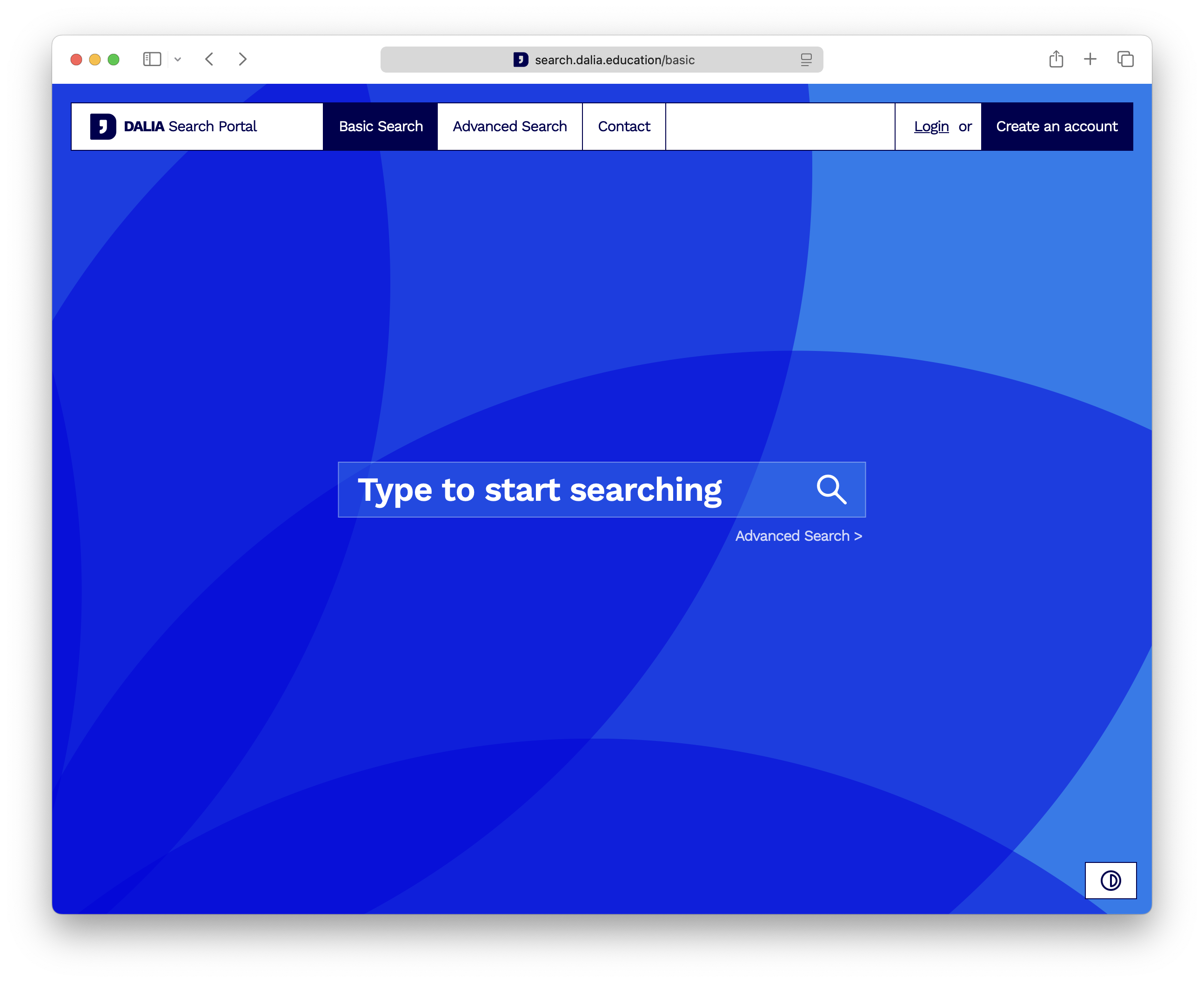}
\hfill
\includegraphics[width=0.49\linewidth]{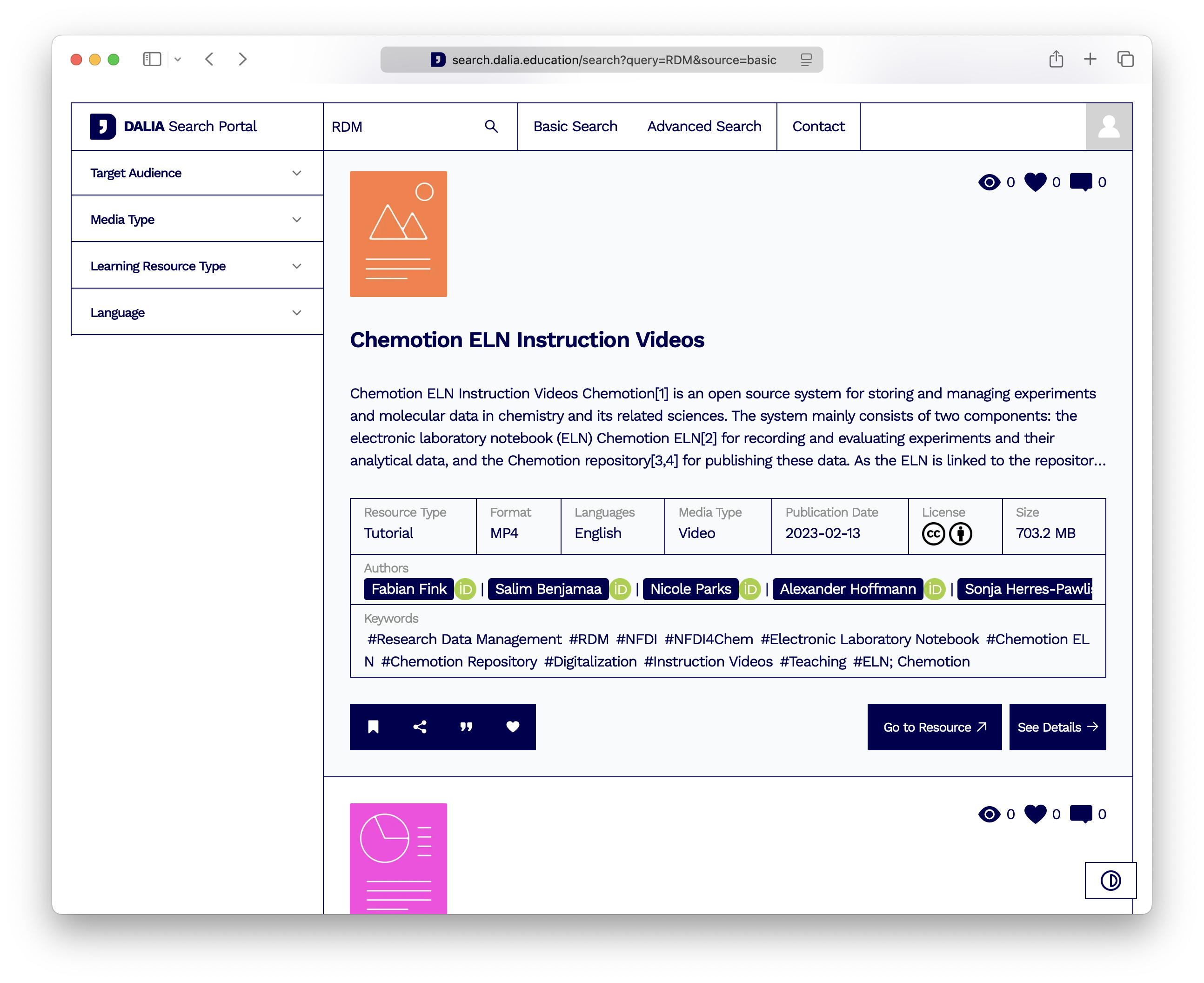}
\caption{The left screenshot shows the DALIA Search Portal, the main entry point for learners and teachers into DALIA. The right screenshot shows the search results page, highlighting the Chemotion ELN Instructional Videos (\url{https://search.dalia.education/items/b37ddf6e-f136-4230-8418-faf18c4c34d2})
}\label{figure-dalia}
\end{figure}

Users of DALIA can perform a faceted search over indexed OERs based on their text fields, then filter based on the classifiers, e.g., enabling a bachelor's level chemist to find OERs specific for their field and educational level (Figure \ref{figure-dalia}). Each OER page shows relevant metadata and links to the external resource and will eventually show suggestions based on users' preferences and recently viewed OERs. Figure \ref{figure-dalia} shows the page for Chemotion ELN Instructional Videos (\url{https://search.dalia.education/items/b37ddf6e-f136-4230-8418-faf18c4c34d2}).

Finally, DALIA implements both a web-based curation interface that enables community-driven submission of new OERs and a spreadsheet-based curation workflow for internally-driven bulk curation (\url{https://git.rwth-aachen.de/dalia/dif-to-dalia-kg}). Community-driven curation supports the longevity of curated resources by enabling small "drive-by" curations by interested community members~\cite{Hoyt2024} and specifically will allow DALIA to best cover the rapidly evolving landscape of OERs in chemistry and other domains.

\section*{Acknowledgments}

The authors thank the DALIA team who worked on the technical implementation of the DALIA platform.

\section*{Funding}

The DALIA project was jointly funded by the the Federal Ministry of Research, Technology, and Space (BMFTR) and the funding measure from the EU's Capacity Building and Resilience Facility with the funding code 16DWWQP07.

\section*{Conflicts of Interest}

The authors declare no conflicts of interest.

\section*{Data Availability}

The video tutorial series presented in the case study is archived on Zenodo at~\cite{Fink2023b}.
The OERs presented in DALIA are stored in version control at \url{https://git.rwth-aachen.de/dalia/dif-to-dalia-kg}. 

\bibliography{main}

\end{document}